\documentclass{article}
\usepackage{spconf,amsmath,graphicx}
\usepackage{multirow}
\usepackage{enumitem}
\usepackage{hyperref}
\setlist{nosep, leftmargin=14pt}

\usepackage{mwe} 

\title{Enhancing 3T Retinotopic Maps Using Diffeomorphic Registration}
 \name{Negar Jalili-Mallak$^{1}$\sthanks{Corresponding author: njalilim@asu.edu} \qquad Yanshuai Tu$^{1}$ \qquad Zhong-Lin Lu$^{2,3,4}$ \qquad Yalin Wang$^{1}$}

 \address{$^{1}$ School of Computing, Informatics, and Decision Systems Engineering,\\ Arizona State University, Tempe, AZ, USA \\
     $^{2}$Division of Arts and Sciences, NYU Shanghai, Shanghai, China.\\
     $^{3}$Center for Neural Science and Department of Psychology, New York University, New York, USA.\\
     $^{4}$NYU-ECNU Institute of Brain and Cognitive Science, NYU Shanghai, Shanghai, China.}

\begin{document}
%
\maketitle
\begin{abstract}
Retinotopic mapping aims to uncover the relationship between visual stimuli on the retina and neural responses on the visual cortical surface. This study advances retinotopic mapping by applying diffeomorphic registration to the 3T NYU retinotopy dataset, encompassing analyze-PRF and mrVista data. Diffeomorphic Registration for Retinotopic Maps (DRRM) quantifies the diffeomorphic condition, ensuring accurate alignment of retinotopic maps without topological violations. Leveraging the Beltrami coefficient and topological condition, DRRM significantly enhances retinotopic map accuracy. Evaluation against existing methods demonstrates DRRM's superiority on various datasets, including 3T and 7T retinotopy data. The application of diffeomorphic registration improves the interpretability of low-quality retinotopic maps, holding promise for clinical applications.
\end{abstract}
\section{Introduction}
\label{sec:intro}

Retinotopic mapping, a fundamental component of visual cognitive neuroscience, helps us understand how the brain processes visual stimuli. Despite its significance, the process relies on functional magnetic resonance imaging (fMRI), which has a low signal–noise ratio (SNR) and relatively low spatial resolution. These limitations impede the creation of precise retinotopic maps. In this study, we explore new methods to enhance the mapping process.

In response to the challenges of retinotopic mapping, we turn our attention to the innovative realm of diffeomorphic registration—a technique known for its remarkable ability to enhance the quality of retinotopic maps \cite{tu2022diffeomorphic, thirion1998image, vercauteren2009diffeomorphic}. Diffeomorphic registration aligns individual subject data with a common template, enhancing the interpretability and utility of retinotopic maps. 

At the forefront of our investigation is DRRM, which introduces a rigorous quantification of the diffeomorphic condition, ensuring that retinotopic maps are accurately aligned through stretching and compressing, without distorting the cortical surface. Leveraging the Beltrami coefficient, DRRM promises to enhance the accuracy of retinotopic mapping while preserving the topological integrity defined in the template.

Our prior research \cite{tu2022diffeomorphic} consistently demonstrated the superior efficacy of the DRRM method across diverse datasets. As we revisit the efficacy of DRRM in this study, our goal is not only to reinforce its superiority over existing registration methods but also to validate its consistent performance on the NYU dataset \cite{himmelberg2021cross}. This involves utilizing 3T imaging and incorporating PRF results derived from methodologies beyond analyze-PRF \cite{kay2013compressive}. This deliberate expansion showcases the versatility and generalizability of DRRM, underscoring its effectiveness in diverse contexts and with varied data sources.

Two prominent methods for population receptive field (pRF) modeling in fMRI studies are analyze-PRF and mrVista. Analyze-PRF, implemented as a command-line-based MATLAB toolbox, is widely utilized in the field \cite{kay2013compressive}. On the other hand, mrVista \href{http://github.com/vistalab}{(http://github.com/vistalab)}, also a MATLAB-based software package, provides a more general-purpose platform that includes pRF modeling among its features. It offers versatility for a broad range of neuroimaging analyses beyond retinotopic mapping.

The importance of handling both analyze-PRF and mrVista results lies in the diversity of tools used by different research groups. As solving pRF is computationally intensive, accommodating both methodologies in the registration pipeline becomes crucial. Our previous work exclusively tested datasets processed using analyze-PRF. However, in this study, we extend our DRRM tool to be compatible with results from both analyze-PRF and mrVista. This compatibility ensures researchers integrate data processed with either tool into our registration pipeline, eliminating the need to recompute pRF solutions and enabling a more efficient, inclusive approach to retinotopic mapping across methodologies.

Furthermore, it's worth noting that while all pRF modeling tools provide essential parameters such as eccentricity, polar angle, and receptive field size, the definition of other parameters, such as the hemodynamic response function (HRF) and receptive field properties, may vary between tools. Different tools may offer additional data that can contribute to more accurate results, depending on how they define these parameters. This emphasizes the need for a flexible registration pipeline capable of handling diverse data outputs from various pRF modeling tools, allowing for a comprehensive analysis of retinotopic maps.

\section{Related Work}
\label{sec:format}

Numerous prior works have significantly contributed to the field of retinotopic map analysis and registration, laying the foundation for the development and application of DRRM. While these efforts have addressed various aspects of retinotopic mapping, none have explicitly incorporated the diffeomorphic condition, a gap successfully addressed by DRRM.

Retinotopic mapping, a cornerstone of vision science, heavily relies on fMRI to delineate the organization of the visual cortex. Approaches such as the populational receptive field (pRF) model \cite{warnking2002fmri, dumoulin2008population} have been pivotal, yet the limitations of fMRI, including low signal-to-noise ratio (SNR) and spatial resolution \cite{vasseur2010fmri}, pose challenges to constructing accurate retinotopic maps.

Various methods, such as smoothing \cite{qiu2006estimating, schira2010modeling}, have been employed to enhance the quality of retinotopic maps, but they often fall short in more intricate visual areas. Diffeomorphic cortical surface registration tools, including FreeSurfer \cite{fischl1999cortical} and Brainsuit \cite{shattuck2002brainsuite, joshi2007surface}, are developed for aligning the brain's cortical surfaces but are unsuitable for retinotopic map registration due to the disparities between retinotopic maps and anatomical surfaces.

Recent advancements like Multimodal Surface Matching \cite{robinson2014msm} improved surface registration but overlook retinotopic coordinates, leaving the diffeomorphic condition unaddressed. Although Benson et al. made significant contributions to retinotopic map alignment \cite{benson2014correction, benson2018bayesian}, their approach lacks explicit consideration of the diffeomorphic condition, crucial for tear-free and distortion-free cortical surface alignment \cite{fischl1999cortical, shattuck2002brainsuite, yeo2009spherical}. Achieving this condition is essential for smooth and invertible registration.

In this context, DRRM emerges as an innovative method \cite{tu2022diffeomorphic},  accommodating analyze-PRF retinotopic mapping method and introducing a new level of versatility and accuracy in retinotopic map alignment. It ensures the diffeomorphic condition, guaranteeing smooth and invertible alignment, and focuses on topological preservation, setting it apart from existing approaches and positioning it as a promising tool for retinotopic mapping research and applications.

The limitations in SNR often lead to noisy data, making it challenging to discern subtle neural responses and accurately define retinotopic boundaries. Constrained spatial resolution can result in inaccuracies, especially in mapping complex retinotopic organizations and fine-scale retinotopic gradients. To address these challenges and enhance the quality of retinotopic mapping,  diffeomorphic registration has gained prominence. Seeking to align individual subject data with a common template while adhering to the diffeomorphic condition, diffeomorphic registration holds the potential to provide more accurate retinotopic maps compared to traditional registration methods.

In this paper, we present a novel application of DRRM, demonstrating its adaptability to the mrVista retinotopic mapping method in addition to the analyze-PRF method. While the original DRRM was designed for analyze-PRF results, our work showcases its extension to incorporate mrVista results, enhancing its versatility across different retinotopic mapping approaches. This adaptability is crucial in addressing the diverse parameters and outcomes associated with these methods. Through the evaluation, we emphasize the continued efficacy of DRRM, underscoring its superiority in achieving diffeomorphic registration. Our results highlight the robustness of DRRM in advancing retinotopic mapping research, expanding its applicability to various datasets, and offering valuable insights for potential clinical applications.

\section{Methods}
\label{sec:pagestyle}

\subsection{Registration}
In our registration process, named Diffeomorphic Registration for Retinotopic Maps (DRRM), we adopt a conformal mapping approach to simplify the 3D-to-2D projection of retinotopic maps, aligning them in the 2D parametric domain \cite{ta2014characterizing, ta2022quantitative}. Following the formulation detailed in the original DRRM paper \cite{tu2022diffeomorphic}, we introduce the concept of a registration energy function ($E_R$), aiming to minimize the differences in retinotopic visual coordinates between subject and template maps while ensuring the smoothness of the registration function. The key feature of DRRM lies in the incorporation of a diffeomorphic condition, quantified using the Beltrami coefficient, to guarantee the topological integrity of the retinotopic maps. This approach is expressed in the mathematical formulation:
\begin{align*}
     \arg \min_f \int \!w \left| v_s(f) - v_T \right|^2 \!+ \lambda_s \left| \nabla f \right|^2 \,\!\! du,
     \text{ s. t. } \left\| \mu_f \right\|_{\infty} < 1 \nonumber
\end{align*}

This represents the 2D registration function, $v_s(f)$ is the visual coordinate of the subject's registered retinotopic map on vertex $i$ and 
$v_T(i)$ is the template's visual coordinate interpreted at position $f_i$. The diffeomorphic condition ensures that $f$ is a quasiconformal map, vital for preserving the topological condition of retinotopic maps. For further details and the numerical solution approach, we refer readers to the original DRRM paper \cite{tu2022diffeomorphic}.

\subsection{NYU retinotopy data}
The NYU Retinotopy Dataset features data from 44 participants recruited from New York University, with a mean age of 28.8 years and an equal gender distribution. Participants underwent a 1 to 1.5-hour scanning session, viewing a population receptive field (pRF) stimulus comprising 100 image patterns within a sweeping bar aperture. Data, acquired on a 3T Siemens MRI scanner, included T1-weighted anatomical and functional echo-planar images. PRF analysis, utilizing mrVista, modeled pRF as a circular 2D-Gaussian with $x$, $y$, and $\sigma$ parameters. The pRF model was implemented on individual participant data, interpolated to fsaverage space, yielding group-average pRF maps \cite{himmelberg2021cross}.

\subsection{Template}
We formulated our template creation by incorporating Benson and Winawer's retinotopic model and integrating the HCP group-average retinotopic map ($T_0 \!=\!(F_{T_0},V_{T_0},v_{T_0},\sigma_{T_0},R^{2}_{T_0})$). Following the DRRM paper \cite{tu2022diffeomorphic}, we transferred the Benson and Winawer retinotopic model from "fsaverage" to "fsLR" space and aligned it with the $T_0$ template using the steps outlined in the "Registration" section. The resulting morphed template, denoted as $T=(F_T,V_T,v_T,\sigma_T,R^{2}_T)$, played a key role as the template within the DRRM framework, facilitating subsequent registration procedures.

\subsection{Performance Evaluation}
To assess the performance of our registration method on the NYU Retinotopy Dataset, we employ indirect metrics due to the lack of ground truth. We measure visual coordinate change ($d\left|v\right|$) for alignment quality, count flipped triangles ($F_{flip}$) to ensure diffeomorphism, and assess goodness of fit to the BOLD time series using RMSE, AIC, and Pearson correlation. Smaller $d\left|v\right|$ and RMSE, along with $F_{flip}=0$, indicate better spatial alignment and topological preservation, while lower AIC values and higher Pearson correlation signify improved fidelity to the BOLD signals.

\section{Results}
\label{sec:typestyle}

\begin{table*}[t]
\caption{Comparison of registration performance metrics (average registration error, RMSE, Pearson correlation, and AIC) for all observers (N = 41) in the NYU Retinotopy Dataset using \textit{mrVista} retinotopic mapping }
\centering
\begin{tabular}{||l|c c c c c c||} 
 \hline
 \textbf{Method} &\textbf{NYU observers} & $d|v|$ & $F_{\text{flip}}$ & \textbf{RMSE } & \textbf{Correlation} $p_c$  & \textbf{AIC } \\ [0.5ex] 
 \hline\hline
 \multirow{2}{*}{DRRM} &Average (L)   &\textbf{3.725}  & \textbf{0} & \textbf{1.076} & \textbf{0.556} & 1045.239 \\ \cline{2-7}
 &Average (R)   &\textbf{3.412}  & \textbf{0} & \textbf{1.045} & \textbf{0.45} & 978.28 \\ 
 \hline\hline
 \multirow{2}{*}{Benson's map} &Average (L)   &3.751  & - & 1.3 & 0.137 & \textbf{918.752} \\ \cline{2-7}
 &Average (R)   &3.849  & - & 1.315 & 0.117 & \textbf{928.588} \\ 
 \hline
\end{tabular}
\label{tab:tab1}
\end{table*}

\begin{table*}[t]
\caption{Performance of DRRM-registered 3T retinotopic maps of all observers (N = 41) in the NYU retinotopy dataset relative to structurally registered retinotopic maps}
\centering
\begin{tabular}{||l|c c c c c c||} 
 \hline
 \textbf{PRF Tool} &\textbf{NYU observers} & $d|v|$ & $F_{\text{flip}}$ & \textbf{RMSE (Raw/Reg)} & \textbf{Correlation} $p_c$ \textbf{(Raw/Reg)} & \textbf{AIC (Raw/Reg)} \\ [0.5ex] 
 \hline\hline
 \multirow{2}{*}{analyze-PRF} &Average (L)   & 2.799 & 0 & 1.149/\textbf{1.129} & 0.33/\textbf{0.352} & 1021.076/\textbf{1006.597} \\ \cline{2-7}
 &Average (R)   & 2.945 & 0 & 1.204/\textbf{1.181} & 0.256/\textbf{0.294} & 1107.628/\textbf{1074.381} \\ 
 \hline
\end{tabular}
\label{tab:tab2}
\end{table*}

In this section, we showcase the results obtained from the application of DRRM to the 3T NYU retinotopy dataset. Our aim is to demonstrate the efficacy and versatility of DRRM in improving the alignment and interpretability of retinotopic maps within the context of this new dataset.

\begin{figure}[t]
  \centering
   \includegraphics[width=1\linewidth]{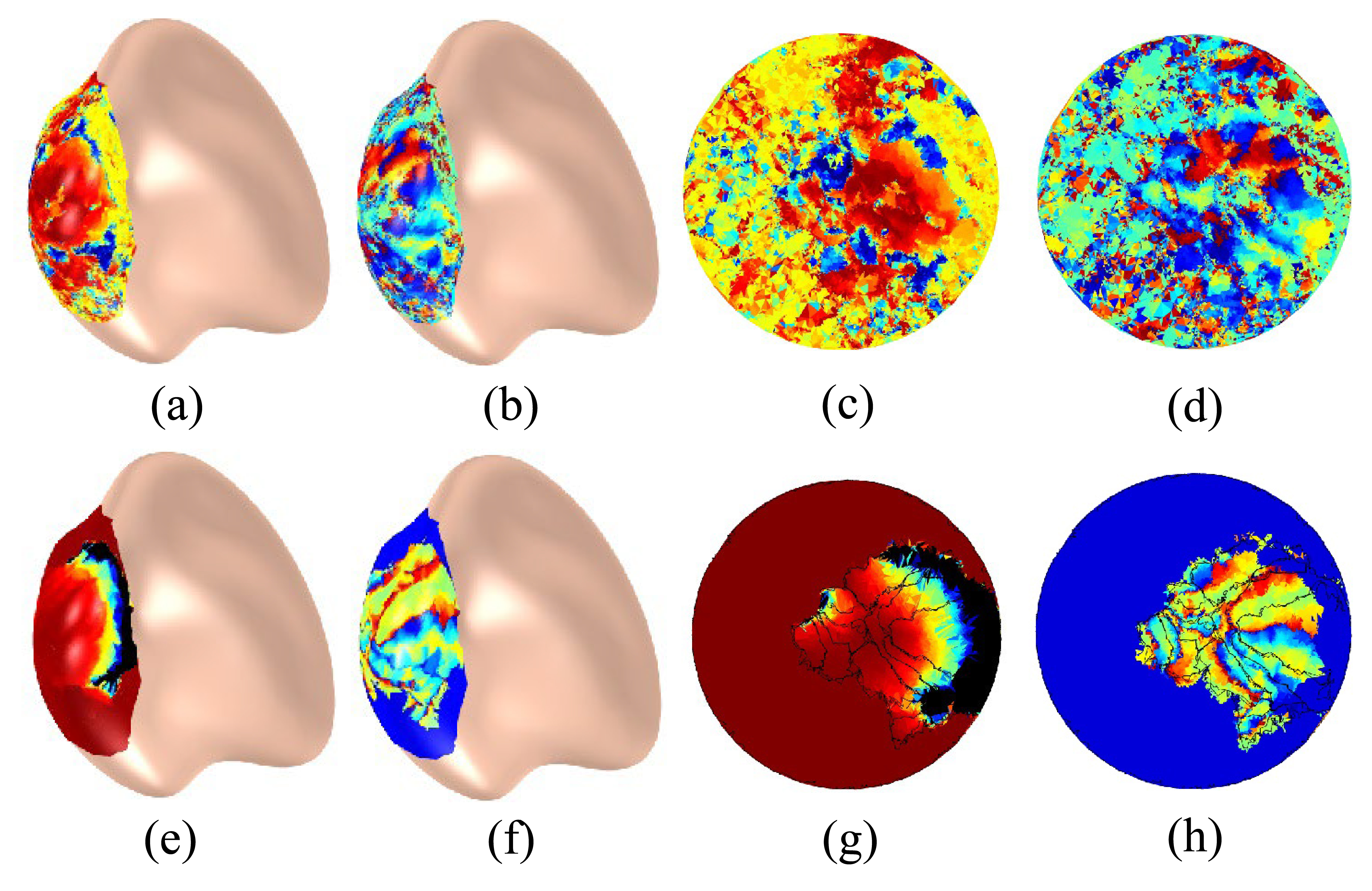}

   \caption{Retinotopic maps for the left hemisphere of one of the observers (subject 19) in the NYU retinotopy dataset \cite{himmelberg2021cross}. Panels a and b depict the eccentricity and polar angle maps, respectively. Panels c and d display the eccentricity and polar angle maps on the disk. Panels e–h exhibit the corresponding registered results.}
   \label{fig:onecol}
\end{figure}

\subsection{Visualization of Retinotopic Maps}
To visually illustrate the impact of DRRM on retinotopic maps, we present results for a representative subject from the NYU retinotopy dataset. The raw retinotopic eccentricity and polar angle results, depicted on the inflated cortical surface for improved visualization, are displayed for the subject's left hemisphere (Fig. \ref{fig:onecol}a and \ref{fig:onecol}b). Subsequently, these results were transferred to the parametric disk (Fig. \ref{fig:onecol}c and \ref{fig:onecol}d). Following the registration process, we updated the visual coordinates for the subject (Fig. \ref{fig:onecol}e-\ref{fig:onecol}h).

\subsection{Quantitative Evaluation}
To assess the performance of DRRM on the NYU retinotopy dataset, we conducted a quantitative evaluation of the registered retinotopic maps,  comparing them with the raw retinotopic maps inferred from FreeSurfer's registration sphere. The absence of flipping triangles in the registered maps demonstrates that DRRM successfully achieves the diffeomorphic condition.

We evaluated the quality of the registered retinotopic maps by examining their fit to the fMRI time series. The results indicate that the DRRM-registered retinotopic maps provide a superior fit to the fMRI time series compared to the structurally registered retinotopic maps. This improvement is evident in the reduced Root Mean Square Error (RMSE) from the DRRM fits, suggesting that the registered visual coordinates provided a better account of the fMRI time series than the original population receptive field (pRF) solutions.(Table \ref{tab:tab1} and \ref{tab:tab2})

Overall, our results underscore the effectiveness of DRRM in enhancing the alignment of retinotopic maps within the NYU retinotopy dataset. This method holds promise for improving the interpretability and utility of retinotopic maps in the context of the 3T fMRI data, extending the application of DRRM beyond its original implementation.

\section{Discussion}
\label{sec:majhead}
In this study, we extended the application of Diffeomorphic Registration for Retinotopic Maps (DRRM) to the 3T NYU retinotopy dataset, building upon the innovative framework introduced in the original DRRM work. Our goal was to assess the adaptability and efficacy of DRRM in the context of a different dataset, analyze-PRF, and mrVista retinotopic mapping methods.

\subsection{Diffeomorphic Registration and Beltrami Coefficient}
Similar to the original DRRM project, our application of
DRRM to the NYU retinotopy dataset emphasized the importance of the Beltrami coefficient in ensuring diffeomorphism
during the registration of individual subject’s retinotopic
maps to a common template. By introducing this coefficient,
we aimed to guarantee not only the smoothness of the registration but also the preservation of the topological condition, where nearby neurons maintain receptive fields at proximal locations on the retina.


\subsection{Preservation of Topological Condition}
A key advantage, as highlighted in the original DRRM work, is the preservation of the topological condition. This preservation ensures that post-registration retinotopic maps are topological, allowing for accurate quantification of essential properties, including cortical magnifications, angle distortions, and boundary differences. Furthermore, DRRM's ability to automatically infer visual area boundaries alleviates the need for tedious manual labeling, emphasizing its efficiency in large-scale analyses.

\subsection{Enhanced Identification of Visual Areas}
The discussion extends to the relevance of diffeomorphic registration in enhancing the identification of visual areas. Structural registration methods, particularly with tools like FreeSurfer, may inadvertently misplace visual boundaries during averaging. DRRM, by emphasizing retinotopic features, addresses this limitation, enabling more accurate identification and delineation of visual areas.


\section{Conclusion}
\label{sec:print}

In this paper, we have applied the Diffeomorphic Registration for Retinotopic Maps (DRRM) framework to the 3T NYU retinotopy dataset, extending its application beyond its original domain. While the original DRRM paper introduced the method, our work demonstrates its adaptability and effectiveness in a different dataset context.

Our application of DRRM to the NYU retinotopy dataset showcases the potential of this method in improving the alignment and interpretability of retinotopic maps. By leveraging the Beltrami coefficient to monitor and preserve the topological condition and employing an iterative algorithm to achieve both the diffeomorphic and topological conditions, we have successfully demonstrated the versatility and accuracy of DRRM in the context of the 3T fMRI data.

Our results highlight the superior accuracy achieved by DRRM, as well as its ability to provide better fits to the Blood Oxygen Level Dependent (BOLD) fMRI time series. This improved alignment and fit to the data underscore the potential of DRRM to enhance retinotopic map analysis.

In conclusion, our work demonstrates that DRRM is a valuable tool that can be successfully applied to different datasets, extending its utility and improving the quality of retinotopic maps in the realm of 3T fMRI data. This adaptability positions DRRM as a promising method for advancing retinotopic map research and applications, addressing the challenges posed by the limitations of fMRI and enhancing the accuracy and interpretability of retinotopic maps.

\section{COMPLIANCE WITH ETHICAL STANDARDS}
\label{sec:ethical}

This research study was conducted retrospectively using human subject data available in open access by \cite{himmelberg2021cross}. Ethical approval was not required as confirmed by the license attached with the open-access data.

\section{Acknowledgments}
\label{sec:acknowledgments}

The work was partially supported by NSF (DMS-1413417 and DMS-1412722) and NIH (R01EY032125 and R01DE030286).


\bibliographystyle{IEEEbib}
\bibliography{refs}

\begin{thebibliography}{10}

\bibitem{tu2022diffeomorphic}
Yanshuai Tu, Xin Li, Zhong-Lin Lu, and Yalin Wang,
\newblock ``Diffeomorphic registration for retinotopic maps of multiple visual regions,''
\newblock {\em Brain Structure and Function}, vol. 227, no. 4, pp. 1507--1522, 2022.

\bibitem{thirion1998image}
J-P Thirion,
\newblock ``Image matching as a diffusion process: an analogy with maxwell's demons,''
\newblock {\em Medical image analysis}, vol. 2, no. 3, pp. 243--260, 1998.

\bibitem{vercauteren2009diffeomorphic}
Tom Vercauteren, Xavier Pennec, Aymeric Perchant, and Nicholas Ayache,
\newblock ``Diffeomorphic demons: Efficient non-parametric image registration,''
\newblock {\em NeuroImage}, vol. 45, no. 1, pp. S61--S72, 2009.

\bibitem{himmelberg2021cross}
Marc~M Himmelberg, Jan~W Kurzawski, Noah~C Benson, Denis~G Pelli, Marisa Carrasco, and Jonathan Winawer,
\newblock ``Cross-dataset reproducibility of human retinotopic maps,''
\newblock {\em Neuroimage}, vol. 244, pp. 118609, 2021.

\bibitem{kay2013compressive}
Kendrick~N Kay, Jonathan Winawer, Aviv Mezer, and Brian~A Wandell,
\newblock ``Compressive spatial summation in human visual cortex,''
\newblock {\em Journal of neurophysiology}, vol. 110, no. 2, pp. 481--494, 2013.

\bibitem{warnking2002fmri}
Jan Warnking, Michel Dojat, Anne Gu{\'e}rin-Dugu{\'e}, Chantal Delon-Martin, Serge Olympieff, Nathalie Richard, Alain Ch{\'e}hikian, and Christoph Segebarth,
\newblock ``fmri retinotopic mapping—step by step,''
\newblock {\em NeuroImage}, vol. 17, no. 4, pp. 1665--1683, 2002.

\bibitem{dumoulin2008population}
Serge~O Dumoulin and Brian~A Wandell,
\newblock ``Population receptive field estimates in human visual cortex,''
\newblock {\em Neuroimage}, vol. 39, no. 2, pp. 647--660, 2008.

\bibitem{vasseur2010fmri}
Flor Vasseur, Chantal Delon-Martin, C{\'e}cile Bordier, Jan Warnking, Laurent Lamalle, Christoph Segebarth, and Michel Dojat,
\newblock ``fmri retinotopic mapping at 3 t: Benefits gained from correcting the spatial distortions due to static field inhomogeneity,''
\newblock {\em Journal of Vision}, vol. 10, no. 12, pp. 30--30, 2010.

\bibitem{qiu2006estimating}
Anqi Qiu, Benjamin~J Rosenau, Adam~S Greenberg, Monica~K Hurdal, Patrick Barta, Steven Yantis, and Michael~I Miller,
\newblock ``Estimating linear cortical magnification in human primary visual cortex via dynamic programming,''
\newblock {\em Neuroimage}, vol. 31, no. 1, pp. 125--138, 2006.

\bibitem{schira2010modeling}
Mark~M Schira, Christopher~W Tyler, Branka Spehar, and Michael Breakspear,
\newblock ``Modeling magnification and anisotropy in the primate foveal confluence,''
\newblock {\em PLoS computational biology}, vol. 6, no. 1, pp. e1000651, 2010.

\bibitem{fischl1999cortical}
Bruce Fischl, Martin~I Sereno, and Anders~M Dale,
\newblock ``Cortical surface-based analysis: Ii: inflation, flattening, and a surface-based coordinate system,''
\newblock {\em Neuroimage}, vol. 9, no. 2, pp. 195--207, 1999.

\bibitem{shattuck2002brainsuite}
David~W Shattuck and Richard~M Leahy,
\newblock ``Brainsuite: an automated cortical surface identification tool,''
\newblock {\em Medical image analysis}, vol. 6, no. 2, pp. 129--142, 2002.

\bibitem{joshi2007surface}
Anand~A Joshi, David~W Shattuck, Paul~M Thompson, and Richard~M Leahy,
\newblock ``Surface-constrained volumetric brain registration using harmonic mappings,''
\newblock {\em IEEE transactions on medical imaging}, vol. 26, no. 12, pp. 1657--1669, 2007.

\bibitem{robinson2014msm}
Emma~C Robinson, Saad Jbabdi, Matthew~F Glasser, Jesper Andersson, Gregory~C Burgess, Michael~P Harms, Stephen~M Smith, David~C Van~Essen, and Mark Jenkinson,
\newblock ``Msm: a new flexible framework for multimodal surface matching,''
\newblock {\em Neuroimage}, vol. 100, pp. 414--426, 2014.

\bibitem{benson2014correction}
Noah~C Benson, Omar~H Butt, David~H Brainard, and Geoffrey~K Aguirre,
\newblock ``Correction of distortion in flattened representations of the cortical surface allows prediction of v1-v3 functional organization from anatomy,''
\newblock {\em PLoS computational biology}, vol. 10, no. 3, pp. e1003538, 2014.

\bibitem{benson2018bayesian}
Noah~C Benson and Jonathan Winawer,
\newblock ``Bayesian analysis of retinotopic maps,''
\newblock {\em elife}, vol. 7, pp. e40224, 2018.

\bibitem{yeo2009spherical}
BT~Thomas Yeo, Mert~R Sabuncu, Tom Vercauteren, Nicholas Ayache, Bruce Fischl, and Polina Golland,
\newblock ``Spherical demons: fast diffeomorphic landmark-free surface registration,''
\newblock {\em IEEE transactions on medical imaging}, vol. 29, no. 3, pp. 650--668, 2009.

\bibitem{ta2014characterizing}
Duyan Ta, Jie Shi, Brian Barton, Alyssa Brewer, Zhong-Lin Lu, and Yalin Wang,
\newblock ``Characterizing human retinotopic mapping with conformal geometry,''
\newblock in {\em SPIE}, 2014.

\bibitem{ta2022quantitative}
Duyan Ta, Yanshuai Tu, Zhong-Lin Lu, and Yalin Wang,
\newblock ``Quantitative characterization of the human retinotopic map based on quasiconformal mapping,''
\newblock {\em Medical image analysis}, vol. 75, pp. 102230, 2022.

\end{thebibliography}

\end{document}